\documentclass[aps,apl,twocolumn,superscriptaddress,showpacs]{revtex4-1}
\usepackage{latexsym}
\usepackage[dvips]{graphicx}
\usepackage[dvips]{color}
\usepackage{graphicx}
\usepackage{amsmath}

\usepackage{ulem}

\begin{document}

\title{Sub-Gap Structure in the Conductance of a Three-Terminal Josephson Junction}

\author{A. H. Pfeffer}
\affiliation{Universit\'e Grenoble Alpes, INAC-SPSMS, F-38000 Grenoble, France}
\affiliation{CEA, INAC-SPSMS, F-38000 Grenoble, France}
\author{J. E. Duvauchelle}
\affiliation{Universit\'e Grenoble Alpes, INAC-SPSMS, F-38000 Grenoble, France}
\affiliation{CEA, INAC-SPSMS, F-38000 Grenoble, France}
\author{H. Courtois}
\affiliation{Universit\'e Grenoble-Alpes, Institut N\'eel, F-38042 Grenoble, France}
\affiliation{CNRS, Institut N\'eel, F-38042 Grenoble, France}
\author{R. M\'elin}
\affiliation{Universit\'e Grenoble-Alpes, Institut N\'eel, F-38042 Grenoble, France}
\affiliation{CNRS, Institut N\'eel, F-38042 Grenoble, France}
\author{D. Feinberg}
\affiliation{Universit\'e Grenoble-Alpes, Institut N\'eel, F-38042 Grenoble, France}
\affiliation{CNRS, Institut N\'eel, F-38042 Grenoble, France}
\author{F. Lefloch}
\affiliation{Universit\'e Grenoble Alpes, INAC-SPSMS, F-38000 Grenoble, France}
\affiliation{CEA, INAC-SPSMS, F-38000 Grenoble, France}
\email[Corresponding author: ]{francois.lefloch@cea.fr}

\begin{abstract}
Three-terminal superconductor (S) - normal metal (N) - superconductor (S) Josephson junctions are investigated. In a geometry where a T-shape normal metal is connected to three superconducting reservoirs, new sub-gap structures appear in the differential resistance for specific combinations of the superconductor chemical potentials. Those correspond to a correlated motion of Cooper pairs within the device that persist well above the Thouless energy and is consistent with the prediction of quartets formed by two entangled Cooper pairs. A simplified nonequilibrium Keldysh Green's function calculation is presented that supports this interpretation.  
\end{abstract}
\maketitle

\section{Introduction}
Josephson effects appear in SNS junctions where two superconductors (S) are electrically coupled through a non-superconducting material (N) \cite{Likharev1991}. The underlying mechanism is the Andreev reflection that converts a Cooper pair in S into two phase-correlated electrons in N \cite{Andreev1966}. 

At zero voltage, the appearance of coherent Andreev bound states leads to a non-dissipative supercurrent through the junction and a mini-gap in the density of states (DOS) of the normal metal. In the case of diffusive junctions and when the length of the normal part $L$ is longer than the superconducting phase coherence length, both the supercurrent and the mini-gap scale with the Thouless energy given by $E_{Th}=\hbar D / L^2$ where $D$ is the diffusive constant of the normal metal.

At finite voltage, the nonequilibrium sub-gap current is governed by Multiple Andreev Reflections (MAR). In this regime, MAR's successively raise a quasiparticle's energy until it reaches the superconducting gap $\Delta$. Due to the superconductor density of states singularity at the gap edge, MAR's lead to a sub-gap structure in the junction differential conductance for $eV=2\Delta/n$ \cite{Klapwijk1982, Octavio1983}, $n$ being an integer. This structure can be observed in diffusive SNS junctions where the diffusion time through the junction is much smaller than the inelastic scattering time. 

In addition to this \mbox{d.c.} sub-gap quasiparticle transport, \mbox{a.c.} Josephson currents also appear in a diffusive SNS junction. However, during the diffusion of an Andreev pair through the junction, phase coherence is maintained only if the energy of the electron or the hole compared to the superconductor chemical potential is smaller than the Thouless energy $E_{Th}$ \cite{Courtois1996}.
The \mbox{a.c.} Josephson current can be indirectly revealed under microwave irradiation. Shapiro steps \cite{Likharev1991} in the \mbox{d.c.} current-voltage characteristics show up when the superconducting phase difference oscillation frequency $2eV/\hbar$ matches the microwave frequency or some multiple of it. The mere existence of Shapiro steps and therefore the \mbox{a.c.} Josephson currents essentially requires a quasi-static superconducting phase difference, \mbox{i.e.} a diffusion time smaller than the inverse of the Josephson frequency $2eV/\hbar$, or equivalently $eV < E_{Th}=\hbar D / L^2$.

More recently, multi-terminal junctions started to be investigated and brought a wealth of new properties, among which several remain to be experimentally uncovered. When two normal conductors are closely connected to a superconducting reservoir, Crossed Andreev Reflections (CAR) can inject two phase-correlated particles, one in each conductor, which amounts to split a Cooper pair into two entangled electrons \cite{Byers1995, Anantram1996, Torres1999, Deutscher2000, Recher2001, Beckmann2004, Cadden-Zimansky2006, Hofstetter2009, Herrmann2010, Das2012}. This only occurs when the distance between the two normal conductors is smaller than the superconducting coherence length.

Another situation is met in mesoscopic three-terminal Josephson junctions in which a single normal conductor is connected to three superconducting contacts \cite{Duhot2009, Houzet2010, Kaviraj2011, Giazotto2011, Galaktionov2012}. The transport properties then depend on two independent (phase or voltage) variables. Therefore, in addition to usual Josephson processes coupling two terminals, new mechanisms are expected that connect all three reservoirs. Several theoretical predictions have been made for such systems \cite{Cuevas2007,Duhot2009,Houzet2010,Freyn2011, Jonckheere2013}. Nonlocal MAR should show up in the so-called incoherent MAR regime where the dwell time exceeds the coherence time \cite{Houzet2010, Chtchelkatchev2010}. On the other hand, the coherent regime where several MAR's can occur within the coherence time is also very interesting. Shapiro-like resonances in the absence of external microwave have been predicted whenever two \mbox{a.c} Josephson frequencies match \cite{Cuevas2007}. On similar grounds, the production of non-local quartets, as pairs of correlated Cooper pairs, has been proposed as a new dissipationless \mbox{d.c.} transport mechanism, which is phase-coherent despite the non-equilibrium conditions \cite{Freyn2011, Jonckheere2013}.
This present work reports on a new experimental study of such phenomena. 

In this article, we report about electronic sub-gap transport in three-terminal Josephson junctions made upon a piece of diffusive normal metal connected to three superconducting reservoirs. The junctions are all phase-coherent as their length is smaller than the single particle phase coherence length $L_\Phi$, and in the long junction regime, e.g. the Thouless energy is much smaller than the superconducting gap. They are also rather symmetric and with a high transparency at every SN interface leading to a large sub-gap Andreev current. Compared to a pair of two-terminal junctions, additional sub-gap structures are observed over a wide voltage range, well above the Thouless energy in a regime where one does not expect the presence of strong \mbox{a.c.} Josephson currents. 

In the following, Section II contains the experimental details and reports the sub-gap anomalies. Section III is devoted to a physical discussion of the possible interpretations. Section IV concludes with perspectives.

\section{The experiment}
\subsection{Samples and measurement process}
\begin{figure}[t]
\includegraphics[width=0.45\textwidth]{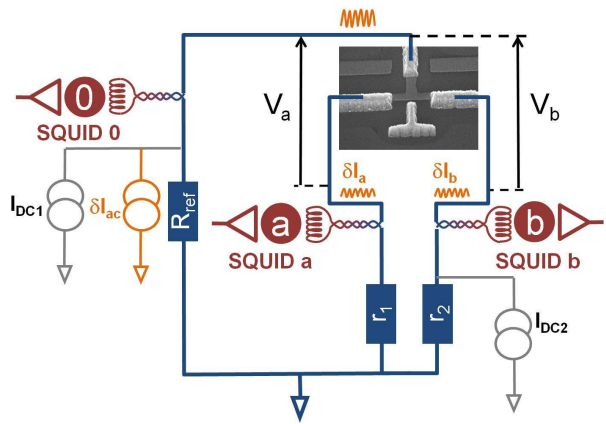}
\caption{Experimental set-up for differential resistance measurements \cite{Pfeffer2012}. The three macroscopic resistors have low resistance values ($\simeq 0.1$ $\Omega$) allowing voltage-biasing the samples. The SEM image shows a three-terminal junction sample with a T-shape geometry.}
\label{Measurement}
\end{figure}

\begin{figure}[t]
\includegraphics[width=0.5\textwidth]{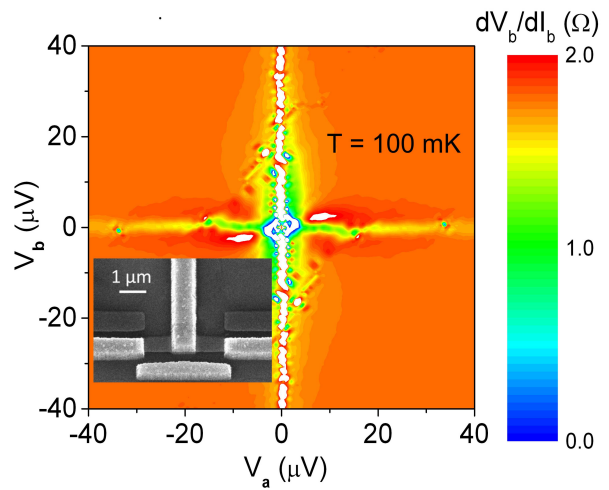}
\caption{Differential resistance $R_{diff,b}$ of a three-terminal device with separated normal metal parts in the ($V_a,V_b$) plane at $T$ = 100 mK. The SEM image represents a sample with such a typical geometry. In this case, only the upper half with $V_b > 0$ has been measured and the graph symmetrized.}
\label{Separated}
\end{figure}

The samples we have studied have been fabricated by shadow mask evaporation technique (see Scanning Electron Microscope SEM images in Fig \ref{Measurement} and \ref{Separated}). Copper and Aluminum were evaporated at different angles through a PMMA/MAA bilayer mask in an ultra-high vacuum chamber. The evaporation of a thin Cu-layer of 50 nm thickness was followed immediately by the evaporation of thick Al electrodes of thickness 500 nm without breaking the vacuum, leading to highly transparent and uniform SN interfaces. The width of the normal metal is about $0.6 \,\mu m$ and its length $L$ is around $1 \,\mu m $. Using a diffusion constant for copper $D = 100 \, cm^2/s$, we get a Thouless energy $E_{Th} = \hbar D/L^2 \simeq 6 \, \mu eV$. This value is confirmed by fitting the temperature dependence of the critical current between two of the superconducting contacts \cite{Dubos2001}. The superconducting Aluminum energy gap is $\Delta = 170 \,\mu eV$ \cite{Hoffmann2004}. 
The diffusion time is $\tau_D=L^2/D \simeq$ 0.1 ns is much smaller than the inelastic time $\tau_{in}\simeq 1 \,ns$ at $100 \, mK$. 

Three-terminal differential resistances were measured using an experimental set-up specially designed to perform highly-sensitive measurements of current average and fluctuations in low-impedance nano-devices at very low temperature \cite{Pfeffer2012}, see Fig. \ref{Measurement}. The experiment operates down to 30 mK and is equipped with 3 commercial SQUIDs as current amplifiers. Each device terminal is connected to the input coil of a SQUID in series with a macroscopic resistor with a low resistance $R_{ref}\simeq r_1\simeq r_2 \simeq 0.09 \,\Omega$. 

The measurement scheme consists in sending an \mbox{a.c.} current modulation $\delta I_{ac} = 1\, \mu A$ on the reference side and recording the current in each branch of the circuit. The differential resistance $R_{diff,a(b)}$ then reads :

\begin{equation}
\label{Eq:Rdiff}
R_{diff,a(b)}= R_{ref}.(\delta I_{ac}-\delta I_{0})/\delta I_{a(b)}-r_{1(2)} 
\end{equation}

where $\delta I_{i}$ is the \mbox{a.c.} current measured in SQUID $i=0,a$ or $b$. For all the samples studied here, $R_{diff,a}$ and $R_{diff,b}$ give the same behavior. In order to explore the non-linear response in the $(V_a,V_b)$ plane, two \mbox{d.c.} current sources were used and the voltage differences $V_a$ and $V_b$ were measured with two room-temperature differential voltage amplifiers. In practice, $I_{DC2}$ is first set to a fixed  value and $I_{DC1}$ ramped with current steps of 1 or 2 $\mu$A. When the ramp is finished, $I_{DC2}$ is increased by a larger current step (typically 20 $\mu$A) and $I_{DC1}$ ramped again. The density of measurement points is therefore not uniform, which explains the dotted features observed in the contour plots.   

\subsection{Results}
Figure \ref{Separated} shows the experimental data measured at $T$ = 100 mK in a sample with separated junctions (see SEM image in Fig. \ref{Separated}). For this sample geometry, the separation holds as the Cu underneath the central electrode, although being continuous, is thin enough that the locally induced gap is that of the superconducting gap $\Delta$ of Aluminum. Only two anomalies corresponding to \mbox{d.c.} Josephson effects at $V_a=0$ and $V_b=0$ are detected. This confirms the absence of multi-terminal effects in the presence of a central electrode with a width ($\sim$ 900 nm) larger than the superconducting coherence length $\xi_s$, as already reported \cite{Kaviraj2011}. Such a device therefore behaves like two independent SNS junctions in parallel. 

Here and in the following, the voltage range was limited to $\pm 40 \,\mu V$ because going beyond would require a \mbox{d.c.} current close to the superconducting electrodes depairing current \cite{Kaviraj2011}. As the investigated voltage range remains well below the superconductor energy gap, the number of multiple Andreev reflections ($\sim 2\Delta /eV$) necessary for a quasiparticle to reach the superconducting gap is more than 8, which would correspond to a total diffusion time much larger than the inelastic scattering time. This defines a strong interaction regime in which MAR cycles are interrupted by inelastic events. In such a bath of thermalized hot quasiparticles carrying an elevated effective temperature, the MAR-induced steps in the energy distribution function are completely washed out \cite{Pierre2001, Hoffmann2004} and hence the sub-gap structures related to the singularity of the DOS at the S/N interface cannot be observed.

\begin{figure}[t]
\includegraphics[width=0.5\textwidth]{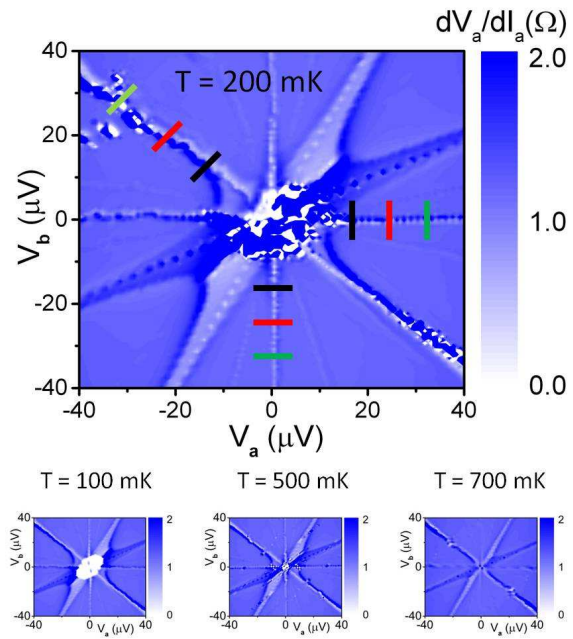}
\caption{Differential resistance $R_{diff,a}$ of a T-shape junction in the ($V_a,V_b$) plane for various temperatures. At $T$ = 200 mK, the data have been measured for the entire voltage range. For the other temperatures, only the upper half with $V_b>0$ has been measured and the graph symmetrized for clarity.}
\label{Tri}
\end{figure}

We have investigated a novel type of three-terminal junction with a T-shape normal conductor connecting three superconducting electrodes $S_0, S_a$ and $S_b$ whose SEM image is shown in Fig. \ref{Measurement}. Here, $S_0$ corresponds to the upper central superconducting electrode, $S_a$ and $S_b$ the left and right superconducting electrodes respectively. The differential resistance $R_{diff,a}$ is shown in Fig. \ref{Tri} at $T=200\,mK$. For this geometry, we expect three Josephson couplings $J_{0a}$, $J_{0b}$ and $J_{ab}$, where the two indexes label the two involved superconducting terminals. In Fig. \ref{Tri}, the couplings $J_{0a}$ and $J_{0b}$ are clearly observed at $V_a = 0$ and $V_b=0$ respectively. As expected from the definition of the differential resistance (Eq. \ref{Eq:Rdiff}), the Josephson coupling $J_{0a}$ appears as a dip in the differential resistance $R_{diff,a}$ whereas the $J_{0b}$ shows up as a peak. We have checked that the opposite behavior is observed when plotting $R_{diff,b}$. 

We can see in Figure \ref{Tri} that the coupling $J_{ab}$ does not show up at $V_b-V_a=0$. In the actual experiment, the \mbox{a.c.}-modulation was sent to the central electrode $S_0$ so that the separation of this current into the two branches $S_a$ and $S_b$ is not sensitive to the coupling $J_{ab}$. We have verified in a similar sample that the latter coupling is indeed revealed when sending the \mbox{a.c.}-modulation through $S_a$ or $S_b$.

In addition to the two \mbox{d.c.} Josephson features discussed above, three other lines are clearly visible at $V_b=-V_a$, $V_b=2V_a$ and $V_b=1/2 V_a$. Notice that in a T-shape geometry, the three superconducting reservoirs are equivalent, meaning that the voltages $V_a$, $V_b$, $V_a-V_b$ are also equivalent. We can thus state that these three lines all originate from the same type of mechanism involving the three superconducting contacts. The observation of this sub-gap structure in the low-bias differential conductance of a three-terminal superconducting hybrid device is the main experimental finding of the present work. 

In a second step, we have studied the temperature dependence of the differential resistance of the T-shape device. The results are plotted at the bottom of Fig. \ref{Tri}. Apart from the central part that is related to the \mbox{d.c.} Josephson effect at very low bias, the sub-gap structure does not evolve much with temperature. All lines are found to be clearly visible up to 700 mK and 40 $\mu V$. This confirms that, in the voltage range under investigation, the electronic temperature is well above the bath temperature \cite{Courtois2008}.

\begin{figure}[h]
\includegraphics[width=0.45\textwidth]{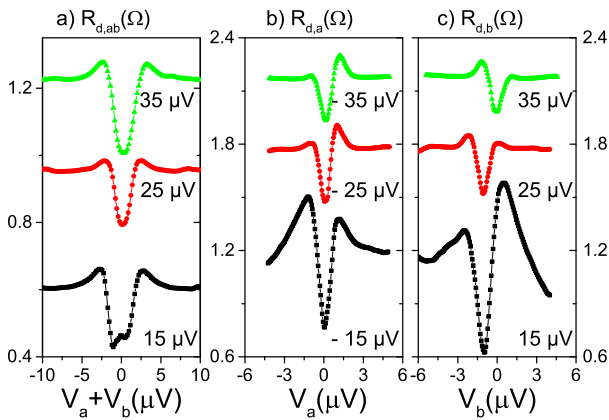}
\caption{Line-traces at various values of the applied voltage of a) the differential resistance of the full sample considering the two branches a,b as being in parallel as a function of the voltage $V_a+V_b$, for several values of  $(V_a-V_b)/2$, b) the differential resistance of branch a vs $V_a$ for various values of $V_b$ and c) the differential resistance of branch b vs $V_b$ for various values of $V_a$. The color code follows that of the lines in Figure \ref{Tri}. The data are shifted for clarity except the lower ones.}
\label{Traces}
\end{figure}

To further investigate these new features, we have plotted some line-traces perpendicular to the $V_b=-V_a$ line (Fig. \ref{Traces}a), to the $V_a=0$ line (Fig. \ref{Traces}b) and to the $V_b=0$ line (Fig. \ref{Traces}c) for various levels of the applied voltages as indicated by the color lines in Figure \ref{Tri}. As expected, the differential resistance $R_{diff,a}$ appears as a \mbox{d.c.} Josephson resonance around $V_a=0$ for any value of $V_b$ (Fig. \ref{Traces}b). The same type of response is observed when plotting $R_{diff,b}$ around $V_b=0$ for any value of $V_a$ (Fig. \ref{Traces}c).  
It turns out that when plotting the overall sample differential resistance recalculated by considering the two branches a and b as being in parallel ($R_{diff,ab}=R_{diff,a}R_{diff,b}/(R_{diff,a}+R_{diff,b})$) as a function of the voltage $V_a+V_b$, the observed profile of the sub-gap structure across the $V_b=-V_a$ line is also in striking resemblance with a Josephson resonance. This observation suggests that the additional anomalies are due to coherent effects involving the three terminals.

Moreover, it is important to notice that the features discussed here are rather robust and constant with respect to the applied voltage. More precisely, as seen in the Fig. \ref{Traces}, those persist at energies well above the Thouless energy.
Therefore, the scheme to explain the additional features seen at non zero $V_a$ and $V_b$ and that involve the three terminals, must also be robust against voltage induced dephasing towards all the branches of the device.   

\section{Interpretation}
\subsection{Synchronization of \mbox{a.c.} Josephson effects}
Sub-gap structures reminding the ones observed here were predicted and observed in the conductance of coupled but separated junctions \cite{Nerenberg1980, Jillie1980}. In this case, two \mbox{a.c.} Josephson currents coexist with frequencies $\nu_a=2eV_a/h$ and $\nu_b=2eV_b/h$. When the two frequencies match, e.g. $V_a=\pm V_b$, down-mixing through the non-linear response of the device can generate \mbox{d.c.} sub-gap structures similar to Shapiro steps. In our experimental scheme with a low resistive environment ($R_{ref}, r_1, r_2 \ll R_n$), this coupling could be obtained through the external circuit. Yet, no anomalies are observed in the sample with separate junctions, despite the fact that both samples have exactly the same circuit environment. In fact, due to the SQUID inductances and the wiring, the external impedances at the Josephson frequency are much larger than the resistances of the bias resistors, preventing any \mbox{a.c} Josephson current to circulate in the external circuit. Therefore, the relevant coupling can only be within the sample itself.

An extended Resistively Shunted Junction (RSJ) model generalizing that of Ref. \cite{Nerenberg1980} could provide a phenomenological description. It involves a triangular Josephson array, shunted by the corresponding normal state resistances, that account for the quasiparticle processes within the N region. With such a model, the observation of strong resonances requires sizeable \mbox{a.c.} Josephson currents, whereas they are known to decrease when the voltages $eV_{a,b}$ exceed the Thouless energy $E_{Th}$ \cite{Volkov1996}. Thus, even if the voltage decrease of \mbox{a.c.} Josephson currents is expected to be progressive, it should definitively lead to a reduction of the resonance for such a variation of $V_{a(b)}$. This is very much in contrast to what is observed in Fig. \ref{Traces}.

In addition to be quantitatively inconsistent, such an RSJ model is only phenomenological. Due to the long coherence time, transport in the present experimental conditions is truly mesoscopic and the explanation of the observations requires a phase-coherent microscopic mechanism taking place in the normal region.

One might consider a more microscopic approach and seek how the possible \mbox{a.c.} Josephson oscillations can synchronize together to yield a constant \mbox{d.c.} component. Such a problem indeed reminds the one considered in Refs. \cite{Argaman1999,Lehnert1999,Dubos2001}. In a clean SNS junction polarized with a voltage $V$,  oscillations with a frequency double of the basic Josephson frequency $\omega=\frac{2eV}{\hbar}$ are generated and show up under microwave irradiation as half-integer Shapiro steps. This was explained by Argaman \cite{Argaman1999} within a semi-phenomenological description in which the both the Andreev levels and their steady state distribution oscillate at the Josephson frequency. The argument applies in the adiabatic regime in which the Thouless energy is much larger than the applied voltage. 

The same argument could be applied to our three-terminal Josephson junctions for which the oscillating \mbox{a.c.} Josephson current at a frequency $\nu_a$ between two of the three terminals could be modulated by oscillation of the distribution function due to the Josephson coupling between two other terminals at a frequency $\nu_b$. In that case, the second harmonic response obtained by Argaman et al. \cite{Argaman1999} in the case of a two terminal SNS junctions, transposes into a response at a frequency $\nu_a +\nu_b$ giving rise to DC features when $\nu_a=-\nu_b$.

Again, the conditions for such a scenario to apply are the same as in Refs. \cite{Argaman1999,Lehnert1999,Dubos2001}, e.g. that the voltage is small enough to allow an adiabatic approximation both in the current components and in the Andreev state distribution. The first one requires that $eV << E_{Th}$, the mini-gap scale, and the second is even more restrictive, $eV << \sqrt{E_{Th}\frac{\hbar}{\tau_in}}$. For instance, in Lehnert et al.'s experiment, the frequency doubling is observed for $eV < 40 \mu eV$, an order of magnitude below the Thouless energy $E_{Th}=350 \,\mu eV$. But in the experiment reported in our work, it is the other way around ! The sub-gap anomalies are observed for $eV$ {\it  above}  $E_{Th}$, up to $8E_{Th}$, only limited by experimental constraints, and without any sign of decay. Thus, though qualitatively appealing, the above mechanism do not provide a good explanation for our experimental results. 

\subsection{The quartet scenario}
\subsubsection{Qualitative description}
The limitation of the synchronization scenario is the voltage induced dephasing suffered by the two electrons of each of the Cooper pairs transferred between two superconducting terminals. Let us instead show that the quartet mechanism, proposed for clean bijunctions \cite{Freyn2011,Jonckheere2013}, can be generalized to a diffusive system and is fully robust against dephasing at voltages much higher than $E_{Th}$. 

The main idea is that two Cooper pairs are transferred in a single and fully energy-conserving quantum process in which the two pairs cross in an entangled way, by exchanging an electron between them. 

\begin{figure}[h]
\includegraphics[width=0.4\textwidth]{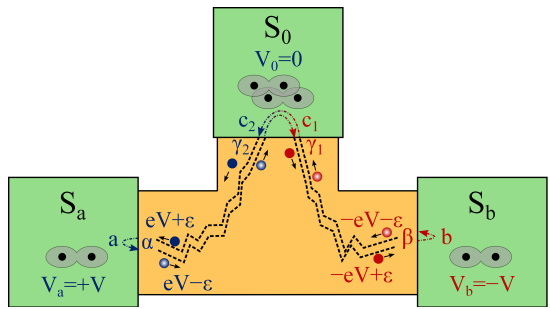}
\caption{Schematic diagram for $Q_0$ quartet production, from $S_0$ to $S_a$ and $S_b$. Two Cooper pairs are split simultaneously at $S_0$ with one electron of each pair propagating towards $S_a$ and $S_b$ where, under the appropriate energy condition ($V_a=-V_b$), they recombine to create two separated Cooper pairs.}
\label{Quartet}
\end{figure}

To describe this mechanism, let us consider a piece of diffusive normal metal $N$ connected to three superconducting reservoirs ($S_0$, $S_a$ and $S_b$) whose potentials are set to $V_0=0$, $V_a=+V$ and $V_b=-V$ respectively, as depicted in Fig. \ref{Quartet}. Two Cooper pairs from $S_0$ can be simultaneous split in N, each of them in two electrons with opposite energies (with respect to the Fermi energy) that we define, without loosing generality, as $\pm (eV+\epsilon)$ and $\pm (eV-\epsilon)$. When these energies are larger than the Thouless energy $E_{Th}$, the two electrons of each pair do not follow the same trajectory. Nevertheless, if the energy $\epsilon$ is small compared to $E_{Th}$, the electron {\it of the first pair} at $eV+\epsilon$ can follow the same phase-coherent trajectory as the electron {\it of the second pair} at $eV -\epsilon$ and for instance, reach $S_a$. 
Since $V_a=+V$, the two particles have relative opposite energies $\pm \epsilon$ and can recombine as a Cooper pair in $S_a$. The same mechanism holds for the two other particles at $S_b$. In the whole process, two Cooper pairs from $S_0$ are split altogether to create two spatially-separated Cooper pairs in $S_a$ and $S_b$, a so-called quartet, named hereafter $Q_0$ as it originates from $S_0$. 

The key point favoring the quartet mechanism is that the coherence of each Andreev pair reaching $S_a$ or $S_b$ can be satisfied at any voltage $V_a = - V_b$, even when $\mid eV_{a(b)}\mid > E_{Th}$. Considering again the four electrons emitted from the two split pairs, two of them have energies $eV \pm \epsilon$ (pair (a)) and the two others $\-eV\pm \epsilon$ (pair (b)) (see Fig. \ref{Quartet}). As the quartet mechanism is a quantum process, the sum of all the possible diffusion probabilities has to be considered altogether. Among those, the situation where pair (a) propagates towards $S_a$ and pair (b) towards $S_b$ is phase coherent and independent of the applied voltage $V$. Indeed, the phase difference accumulated by the pair (a) (pair (b)) scales as $\epsilon\tau_{Da(b)}/\hbar$ where $\tau_{Da(b)}$ is the diffusion time from $S_0$ to $S_a (S_b)$. The quartet mode is therefore a fully coherent d.c. process taking place in the mesoscopic $N$ region and involving four Andreev reflections. 

This is very different from the scheme where two electrons of a single Cooper pair propagate towards $S_a$ or $S_b$. In that case, the two electrons of a single pair have energies $\pm (eV+\epsilon)$ or $\pm (eV-\epsilon)$ and the accumulated phase difference scales with $eV/E_{Th}$. The effect of such trajectories has, therefore, a vanishing contribution to the electronic transport when $eV \gg E_{Th}$.   

Let us note that the quartet response bears some resemblance with MAR's \cite{Houzet2010}, with two important differences. First, the total energy balance of the process is zero, and second, it does not lead to quasiparticle transport above the superconducting gap. 

In this quartet description, the line at $V_b=-V_a$ corresponds to the production of quartets $Q_0$, whereas the line at $V_b=2V_a$ ($V_b=1/2 V_a$) originates from quartets $Q_a$ ($Q_b$) produced in $S_a$ ($S_b$) towards $S_0$ and $S_b$ ($S_0$ and $S_a$). At lowest order, the quartet mechanism requires only four Andreev reflections, much less than needed in the same voltage range for a quasiparticle to reach the gap edge in a usual MAR process, which makes the quartet mechanism much more robust to inelastic collisions.

\subsubsection{Sketch of the microscopic calculation}
The above arguments can be substantiated by a microscopic calculation (Appendix A), valid under the separation of energy scales $E_{Th} < eV  < \Delta$. One uses non-equilibrium Keldysh Green's functions and performs a lowest order calculation in tunnel amplitudes at the different SN interfaces, inspired by Ref. \cite{Cuevas1996}. The quartet current is calculated using a Hamiltonian formalism and an essential step is averaging over disorder. 

The current appears as a sum of contributions, each being a product of six propagating amplitudes associated to the diagrammatic lines in Fig. \ref{Quartet} (Eq. \ref{eq:diff-proba1} - \ref{eq:diff-proba2}). As a classical procedure in the treatment of diffusion in metals and in an SNS junction \cite{Akkermans2007}, disorder averaging takes advantage of the energy separation of these lines. In fact, two lines are correlated by disorder provided their energies are closer than $E_{Th}$. It results that the averaged product of six amplitudes can be decoupled into three factors. One corresponds to the diffusion of an Andreev pair from $S_0$ to $S_a$, at energies $eV \pm \varepsilon$, with $\varepsilon < E_{Th} < eV$, another one to the diffusion of an Andreev pair from $S_0$ to $S_b$, at energies $-eV \pm \varepsilon$ and the third to the anomalous diffusion within $S_0$ that achieves Andreev reflection at energies $eV$, $-eV$ \cite{Feinberg2003}. 

The principle of the above calculation can be benchmarked on the simpler case of an $SNS$ junction at equilibrium for which 
the coherent {\it pair} current is proportional to the {\it single particle} conductance $G_N$ times the coherent energy window given by the Thouless energy. This leads to the known scaling for the critical current $e I_c \propto G_N E_{Th}$ \cite{Dubos2001}.

The main result of our calculation detailed in the Appendix, is to show that the coherent {\it quartet} current has a similar form and is given by the {\it two-particle} CAR conductance times the same energy window $E_{Th}$. It follows a scaling given by :
\begin{equation}
\label{eq:IQ}
eI_Q \sim -G_{CAR} E_{Th}
\end{equation}
The minus sign comes from the exchange and recombination process \cite{Jonckheere2013}. The conductance $G_{CAR}$ is the Crossed Andreev conductance of a $N_aNS_0NN_b$ structure in which the electrodes $S_a$ and $S_b$ are in the normal state and at voltages $\pm V$. 

The CAR conductance can then be evaluated (see Appendix) and recast as :

\begin{equation}
G_{CAR} \sim \frac{G_{Na}G_{Nb}}{G_{0}(\xi_s)}
,
\end{equation}
where $G_{Na,b}$ is the conductance within each normal branch of the bijunction, and $G_0(\xi_s)$ is the normal-state conductance of a region of size $\xi$ of the superconductor $S_0$.
This calculation shows that the ratio between the quartet maximum current at a bias $V$ and the single junction critical current at zero bias is $I_Q/I_c(0) \sim G_{CAR}/G_N \sim G_N/G_0(\xi_s)$, which is not necessarily small. Based on measured sample parameters, we estimate this ratio to $\sim 0.1-0.5$, in fair agreement with the experiment. Notice that if $eV<<\Delta$, $G_{CAR}$ thus $I_{Q,max}$ does not decrease with $V$, in agreement with the present experiment.

\section{conclusion}
In conclusion, we reported about new sub-gap structures in the differential conductance of a metallic nanostructure with three superconducting reservoirs, a so-called bijunction. The existence of such anomalies well above the Thouless energy points towards a new and fully coherent mechanism, different from the synchronization of separated Josephson junctions, or any mesoscopic generalization of such a process. Our results are consistent with the production of non-local quartets, as a resonant pair of Cooper pairs, splitting and recombining within the N region. Therefore, our results provide a convincing experimental evidence for (double) crossed Andreev reflections in metallic superconducting / normal metal hybrid three-terminal nanostructures with a signature in the electronic response at low temperature much larger than in metallic Cooper pair splitters using only one superconducting reservoir. 

The quartet mechanism carries intrinsic four-particle entanglement, generalizing two-fermion entanglement from CAR \cite{Recher2001} that could be exploited if adding more degrees of freedom such as in quantum dots \cite{Freyn2011}. 
More refined probes are necessary to quantitatively study the correlated pair transport involving quartets, as well as possible other regimes, not evidenced in the present experiments, like the low-voltage adiabatic transport. An useful tool is to couple the bijunction to microwaves and study the Shapiro steps coming from deviations from the resonant situation $V_a=-V_b=V$. 

This work has been partially funded by the French Research National Agency, ANR-NanoQuartet (ANR12BS1000701). We acknowledge the Nanoscience Foundation for the PhD grant of A. H. Pfeffer, the NanoFab facility at Institut N\'eel-CNRS for sample fabrication and fruitful discussions with B. Dou\c{c}ot and C. Padurariu. 

\newpage

\appendix
\section{Analytical calculation of the quartet current}
The superconductors $S_{0,a,b}$ are described by the mean-field BCS
Hamiltonian, with identical gaps $\Delta$ and phases $\varphi_{0}=0$, $\varphi_{a}$, $\varphi_{b}$. To simplify, all materials $S_i,N$ are taken with the same bandwidth $w$, and they are connected by a hopping parameter $\tau$, related to the interface transparency $T$ by 
$T=\frac{4\tau^2/w^2}{(1+\tau^2/w^2)^2}$. 
In Nambu notations, the
hopping amplitudes take the form ($i=0,a,b$, and $\alpha$ denoting the position on the interface)
\begin{eqnarray}
\label{eq:Nambu-hopping-11}
\hat{\Sigma}_{i,\alpha}(t)&=&\tau \left(\begin{array}{cc}
e^{ieV_i t}&0\\0&-e^{-ieV_i t}\end{array}\right)
.
\end{eqnarray}
The local advanced Green's functions in the superconductors are as follows in thre frequency domain ($\omega_{\eta}=\omega-i\eta$):
\begin{eqnarray}
\hat{g}_{i,i}^A(\omega)&=&\frac{1}{w\sqrt{\Delta^2-\omega_{\eta}^2}}
\left(\begin{array}{cc}-\omega_{\eta} &
  \Delta\ e^{i\varphi_i}\\ \Delta e^{-i\varphi_i}&
  -\omega_{\eta} \end{array}\right)
,
\end{eqnarray}
The retarded Green's functions are obtained by changing $\eta$ into $-\eta$ in the above expression.
The choice of the
gauge is such that the time-dependence of the phases $2eV_{a(b)} t/\hbar$ are
included in the Nambu hopping amplitudes $\hat{\Sigma}_{a(b)}$ (with
$\hbar=1$). The phases $\varphi_{a(b)}$ at the origin of time are included in the
off-diagonal components of the Nambu Green's functions. The local advanced Green's functions are $\hat{g}^A(\omega)\sim i \pi \rho_N$ in the normal metal, where $\rho_N$ is the local density of states of the normal metal. 

The current at some point $a$ of the interface of the superconductor $S_a$ is given by
\begin{align}
I_{a}(t)=\frac{2e}{h}{\cal R}e\left[\hat{\Sigma}_{a,\alpha}(t)
  \hat{G}^{(+-),11}_{\alpha,a}(t,t)- \right. \nonumber\\
   \hat{\Sigma}_{a,\alpha}(t)  \left.
  \hat{G}^{(+-),22}_{\alpha,a}(t,t)\right]
\end{align}
where $\hat{G}^{(+-),11}_{\alpha,a}(t,t)$ (resp. $\hat{G}^{(+-),22}_{\alpha,a}(t,t)$) is the electron (resp. hole) Keldysh Green's function at point $a$. 

Together with $\hat{G}^{(R,A)}$, $\hat{G}^{(+-)}_{\alpha,a}(t,t)$ obeys a Dyson equation which allows to calculate the current as a product of Green's functions propagating electrons (holes) in the normal or superconducting regions, and hopping self-energies 
$\hat{\Sigma}$ at the interfaces. Stationarity allows to Fourier transform the time quantities and calculate the current contributions as a sum over the Fourier components $\hat{G}(\omega_n)$ with $\omega_n=\omega+neV$. Specifying to the voltages $V_a=V, V_b=-V$, the self-energies $\hat{\Sigma}(\omega)$ connect  Green's functions with indices $n$ differing by $\pm1$. 

The quartet diagram on Fig. \ref{Quartet} takes a typical chain form, starting at the $S_0-N$ interface (with the frequency arguments omitted):
{\small\begin{eqnarray}
\label{chain}
\nonumber
({\cal A})_Q = \Sigma_{c_1,\gamma_1}^{11/00}
g_{\gamma_1,\beta}^{11/00}
\Sigma_{\beta,b}^{11/01}
g_{bb}^{12/11}
\Sigma_{b,\beta}^{22/12}
g_{\beta,\gamma_1}^{22/22}\nonumber\\
\Sigma_{\gamma_1,c_1}^{22/22}
g_{c_1,c_2}^{21/22}
\Sigma_{c_2,\gamma_2}^{11/22}
g_{\gamma_2,\alpha}^{11/22}
\Sigma_{\alpha,a}^{11/21}
g_{aa}^{12/11}\nonumber\\
\Sigma_{a,\alpha}^{22/10}
g_{\alpha,\gamma_2}^{22/00}
\Sigma_{\gamma_2,c_2}^{22/00}
g_{0_2,c_1}^{21/00}
.
\end{eqnarray}}
The first two upper labels correspond to Nambu matrix notation and the second two to the
harmonics $(n,n')$ of half the Josephson frequency $\omega_0=\frac{2eV}{\hbar}$. The advanced, retarded
and Keldysh labels have to be inserted in this expression, resulting in eight different
terms. Next, each of the eight terms is evaluated. The final expression for $(A_Q)$ is
as follows:
\begin{eqnarray}
\label{amplitudes}
\nonumber
({\cal A})_Q = n_F(\omega-eV)\tau^8 \\
\left \{
g^{A}_{aa}
g^A_{\alpha,\gamma_2}
g^{A}_{\gamma_2,\gamma_1}
g^A_{\gamma_1,\beta}
g^{A}_{bb}
g^A_{\beta,\gamma_1}
g^{A}_{\gamma_1,\gamma_2}
g^A_{\gamma_2,\alpha} 
- A \leftrightarrow R\right \} 
\end{eqnarray}
where $A\leftrightarrow R$ means that ``advanced'' and ``retarded'' have been
interchanged. The unperturbed Green's functions $g^A_{ij}$ represent the amplitudes for electron and hole propagation and they are evaluated at the appropriate energies
$\pm eV \pm \varepsilon$ shown on Fig.\ref{Quartet}. Those energies correspond to the
transitions between $n$ and $n'$ indices (see Eq. (\ref{chain})) induced by the hopping matrix elements. A
summation over the labels $\alpha$ ($\beta$) and $\gamma$ at the
interfaces has to be carried out. This procedure is justified to describe
extended contacts at lowest order in the tunnel amplitudes. 
As far as the applied voltages are small
enough compared to the gap, the energy dependence of the Green's
functions can be discarded.

The next step is to perform disorder averaging. A contribution such as $({\cal A})_Q$ should be replaced by
its average $\langle\langle{({\cal A})_Q}\rangle\rangle$ over disorder in the $N$ region and in the superconductors. 
Expression (\ref{amplitudes}) contains several amplitudes, matrices in $S_i$, numbers in $N$. 
First, $g^A_{aa}$, $g^A_{bb}$ yielding density of states factors in $S_a$ and $S_b$. Second, the product $(g^{A}_{\gamma_2,\gamma_1}g^{A}_{\gamma_1,\gamma_2})$ of amplitudes in $S_0$, at energies close to $eV$ (electrons) and $-eV$ (holes) that can be averaged separately. 
It describes the anomalous diffusion of a quasiparticle within $S_0$, yielding Andreev-reflection at the $NS_0$ interface \cite{Feinberg2003}. Third, the product $(g^A_{\alpha,\gamma_2}g^A_{\gamma_2,\alpha}g^A_{\gamma_1,\beta}g^A_{\beta,\gamma_1})$ of amplitudes in $N$. The two first amplitudes contribute at energies $eV \pm \varepsilon$ and the two others at energies $-eV \pm \varepsilon$. 
We assume that $eV > E_{Th}>\varepsilon$, and use the fact that the coherence between electron and hole trajectories is limited by the Thouless energy. 
Then it is justified to decouple $\langle\langle{g^A_{\alpha,\gamma_2}g^A_{\gamma_1,\beta}g^A_{\beta,\gamma_1}g^A_{\gamma_2,\alpha}}\rangle\rangle \simeq \langle\langle{g^A_{\alpha,\gamma_2}g^A_{\gamma_2,\alpha}}\rangle\rangle\langle\langle{(g^A_{\gamma_1,\beta}g^A_{\beta,\gamma_1}}\rangle\rangle$. 
This amounts to separately averaging the diffusive trajectories connecting $S_a$ to $S_0$, and $S_b$ to $S_0$, relying on energy rather than spatial separation. Then one obtains :
\begin{eqnarray}
\label{eq:decouplage}
\nonumber
\langle\langle({\cal A})_Q\rangle\rangle
= 2n_F(\varepsilon-eV)(\pi \rho_N)^2 \tau^8 \\
\left \{P_{\beta,\gamma_1} 
{\tilde P}_{c_1,c_2}
P_{\gamma_2,\alpha} 
\right \} \sin(\varphi_a+\varphi_b).
\end{eqnarray}
with
\begin{eqnarray}
\label{eq:diff-proba1}
P_{\beta,\gamma_1}&\equiv& \langle\langle{g^{11}_{\gamma_1,\beta}(\epsilon-eV)
g^{22}_{\beta,\gamma_1}(\epsilon+eV)} 
\rangle\rangle\\
P_{\gamma_2,\alpha}&\equiv& \langle\langle{g^{11}_{\gamma_2,\alpha}(\epsilon+eV)
g^{22}_{\alpha,\gamma_2}(\epsilon-eV)}
\rangle\rangle\\
{\tilde P}_{c_1,c_2}&\equiv&
\langle\langle g^{12}_{c_1,c_2}(\epsilon-eV)
 {g}^{21}_{c_2,c_1}(\epsilon+eV) \rangle\rangle .
\label{eq:diff-proba2}
\end{eqnarray}
and where the characteristic phase dependence of the quartet mode stems from the
four involved Andreev reflections, one at $S_a$, one at $S_b$ and two at
$S_0$. A product of three probabilities appear : $P_{\gamma_2,\alpha}$ for electron-hole (Andreev pair) diffusion from $S_a$ to $S_0$, ${\tilde P}_{c_1,c_2}$ for the anomalous diffusion inside $S_0$, and $P_{\beta,\gamma_1}$ for the Andreev pair diffusion from $S_0$ to $S_b$. 
${\tilde P}_{c_1,c_2}$ tracks the probability of two Andreev reflections at $S_0$. 
The Andreev pair diffusion modes $P_{ij}=P(R_{ij},\omega)$ showing out in Eq. \ref{eq:decouplage}  are obtained by a summation of the ladder
diagrams \cite{Akkermans2007}, standard in the diffusion problem. 

The diffusion probability on a distance $R$ is $P_0(R,\omega,V)=
\langle\langle g^{11,A}(R,\omega-eV) g^{22,A}(R,\omega+eV) \rangle\rangle$ and its space Fourier transform is proportional to
\begin{equation}
\label{eq:P0}
P_0({\bf q},\omega,V)\sim \frac{1}
{w(i\omega+D q^2)}.
\end{equation}
Importantly, $P_0({\bf q},\omega,V)$ has no dependence on $V$, in the $V << \Delta$ limit. Most importantly, the quartet current appears even if the voltage is larger than
the Thouless energy.

The above principle for the quartet current calculation can be benchmarked on the simpler case of an $SNS$ junction at equilibrium.  
One considers the DC-Josephson current in a $S_aINIS_b$ junction, and evaluate it on the same line as above, by a expansion of the current to fourth order in the transparencies. Then the Fourier
transform of the diffusion probability associated to the Andreev pair modes
$\langle\langle g^{11,A}_{\gamma_1,\beta}(\omega-eV) g^{22,A}_{\beta,\gamma_1}
(\omega+eV) \rangle\rangle$ in $N$ is also given by $P_0({\bf q},\omega,V)$.
For comparison, in a $N_aININ_b$ junction with the mode $\langle\langle
g^{11,A}_{\gamma_1,\beta}(\omega-eV) g^{11,R}_{\beta,\gamma_1} (\omega+eV)
\rangle\rangle$, the diffusion probability is $P_0({\bf q},0,V)$, thus without
the $i\omega$ factor in the denominator of Eq.~(\ref{eq:P0}).

Depending on the diffusion taking place in N or S, this results after integration over $q$:
\begin{eqnarray}
\label{diffusion}
P_{0N}(R,\varepsilon)&\sim& 2\pi \rho_N \frac{1}{2DR} exp\left(-\sqrt{\frac{\varepsilon}{E_{Th}}}\right)cos\left(\sqrt{\frac{\varepsilon}{E_{Th}}}\right)\nonumber\\
{\tilde P}_{0S}(R,\varepsilon)&\sim& 2\pi \rho_N \frac{1}{2DR} exp\left(-\frac{R}{\xi}\right)
\end{eqnarray}
where the Thouless energy for a junction of length $R$ is $E_{Th}=\frac{\hbar D}{R^2}$. Notice that ${\tilde P}_{0S}$ decays on the effective coherence length in $S_0$ (taking into account disorder) \cite{Feinberg2003}.

The quartet current calculation follows from equations (A3-A6).
\begin{eqnarray}
\label{eq:IQ0}
\nonumber
I_{tot}\sim \frac{2e}{h} {\cal N} \frac{\tau^8}{w^2} \int{d\omega}
 P_{\beta,\gamma_1} (\omega)
P_{\gamma_2,\alpha} (\omega)\\
\int_{{\cal S}_c}\frac{d^2r}{\xi^2}  {\tilde P}_{c_1,c_2} (\omega) (\delta{\cal V})^3\sin(\varphi_a+\varphi_b)
\end{eqnarray}
where ${\cal N}$ is the average number of channels, due to integration on one of the interfaces of $S_{a,b}$. The integration volume $\delta{\cal V}\sim \lambda_F l_e^2$ accounts for the absorbing boundary conditions for diffusion in the reservoirs \cite{Akkermans2007} ($\lambda_F$ is the Fermi wavelength). Integration over the surface of $S_0$ accounts for the range $\xi$ of the Andreev reflection and yields a total factor $\frac{\xi}{wl_e}$ for the integrated Andreev probability in $S_0$.  Integration over frequency yields the factor $E_{Th}$, each diffusion probability contributes by a factor $\frac{l_e}{wL}$. One finally obtains: 
\begin{equation}
\label{eq:IQ2}
eI_Q \sim -G_{CAR} E_{Th} \sin(\varphi_a+\varphi_b),
\end{equation}
A yet unknown prefactor has to be added in Eq.~(\ref{eq:IQ}), which is expected to be of the same order as that involved in the case of a $SNS$ junction.

The conductance $G_{CAR}$ refers to the Crossed AndreevReflection (CAR) in a $N_aNS_0NN_b$ hybrid structure. The maximum quartet current is thus naturally obtained by multiplying the CAR conductance by the Thouless energy that sets the coherence of Andreev pairs on both branches a,b of the bi-junction.

The CAR conductance is evaluated from above:
\begin{equation}
G_{CAR}\sim\frac{2e^2}{h}  {\cal N} \left(\frac{\tau}{w}\right)^8 \left(\frac{l_e}{L}\right)^2 \frac{\xi}{l_e},
\end{equation}
where the ratio $\frac{\tau}{w}$ is taken from the rather good experimental conductance $T\sim 0.3$.

%

\end{document}